\documentclass[12pt]{article}
\usepackage{graphicx}
\usepackage{epsfig}
\usepackage{epstopdf}
\DeclareGraphicsExtensions{.pdf,.eps,.png,.jpg,.mps}

\def\lsim{\mathrel{\rlap{\lower3pt\hbox{\hskip0pt$\sim$}}
     \raise1pt\hbox{$<$}}}         
\def\gsim{\mathrel{\rlap{\lower4pt\hbox{\hskip1pt$\sim$}}
     \raise1pt\hbox{$>$}}}         

\usepackage{amsmath}
\usepackage{amsfonts}

\begin{document}
\begin{titlepage}

\centerline{\Large \bf Path Integral and Asset Pricing}
\medskip

\centerline{Zura Kakushadze$^\S$$^\dag$\footnote{\, Zura Kakushadze, Ph.D., is the President of Quantigic$^\circledR$ Solutions LLC, and a Full Professor at Free University of Tbilisi. Email: \tt zura@quantigic.com}}
\bigskip

\centerline{\em $^\S$ Quantigic$^\circledR$ Solutions LLC}
\centerline{\em 1127 High Ridge Road \#135, Stamford, CT 06905\,\,\footnote{\, DISCLAIMER: This address is used by the corresponding author for no
purpose other than to indicate his professional affiliation as is customary in
publications. In particular, the contents of this paper
are not intended as an investment, legal, tax or any other such advice,
and in no way represent views of Quantigic® Solutions LLC,
the website \underline{www.quantigic.com} or any of their other affiliates.
}}
\centerline{\em $^\dag$ Free University of Tbilisi, Business School \& School of Physics}
\centerline{\em 240, David Agmashenebeli Alley, Tbilisi, 0159, Georgia}
\medskip
\centerline{(October 6, 2014; revised: February 20, 2015)}

\bigskip
\medskip

\begin{abstract}
{}We give a pragmatic/pedagogical discussion of using Euclidean path integral in asset pricing. We then illustrate the path integral approach on short-rate models. By understanding the change of path integral measure in the Vasicek/Hull-White model, we can apply the same techniques to ``less-tractable" models such as the Black-Karasinski model. We give explicit formulas for computing the bond pricing function in such models in the analog of quantum mechanical ``semiclassical" approximation. We also outline how to apply perturbative quantum mechanical techniques beyond the ``semiclassical" approximation, which are facilitated by Feynman diagrams.
\end{abstract}
\medskip
\end{titlepage}

\newpage

\section{Introduction}

{}In his seminal paper on path integral formulation of quantum mechanics, Feynman (1948) humbly states: ``The formulation is mathematically equivalent to the more usual formulations. There are, therefore, no fundamentally new results. However, there is a pleasure in recognizing old things from a new point of view. Also, there are problems for which the new point of view offers a distinct advantage." Feynman was referring to his path integral formulation of quantum mechanics he described in that paper in relation to the existing equivalent formulations, Schr\"odinger's wave equation and Heisenberg's matrix mechanics. Subsequently, he applied path integral to Quantum Electrodynamics and developed the Feynman diagram techniques (Feynman, 1949), which have been used to compute various experimentally measured quantities in quantum field theory with astounding precision.

{}That the Euclidean version of Feynman's path integral can be applied in finance, including in asset pricing problems, has been known for quite some time. Just as in quantum mechanics, path integral in finance is neither a panacea, nor is it intended to yield ``fundamentally new results". Instead, again, just in quantum mechanics (and quantum field theory), it is an equivalent formulation, which in some cases provides intuitive clarity and insight into old problems. Thus, where stochastic differential equations and pricing PDEs just happen to be cumbersome to use or even difficult to write down, path integral can sometimes provide a clearer view of a pathway toward a possible solution. It is with this understanding that in these notes we attempt to discuss path integral in the context of asset pricing.

{}We start with classical mechanics (Section \ref{sec.2}) and then discuss Feynman's path integral formulation of quantum mechanics (Section \ref{sec.3}).\footnote{\, Some readers may wish to skip Sections \ref{sec.2} and \ref{sec.3}.} We do not give a derivation of Feynman's path integral -- our goal is path integral in asset pricing. To this end, we then discuss Euclidean path integral (Subsection \ref{sub.euclidean}), which is what is relevant in asset pricing (Section \ref{sec.4}). In the asset pricing context we discuss the analog of the semiclassical approximation, which in quantum mechanics amounts to keeping the leading quantum correction, whereas in asset pricing it has the meaning of small ``volatility" approximation (discussed in Section \ref{sec.7}). In the pricing PDE language, this is the WKB approximation. Path integral in this approximation is Gaussian. Going beyond this ``semiclassical" approximation amounts to doing perturbation theory, which is well understood in quantum mechanics as well as Euclidean path integral and is greatly facilitated by the Feynman diagram techniques.

{}There are various applications of path integral in asset pricing, including option pricing and interest rate products. A natural application is to bond pricing in short-rate models. In Section \ref{sec.5}, to illustrate usage of path integral techniques, we give a two-line derivation of the bond pricing function in the Vasicek/Hull-White model. We then take a harder route and derive the same result (in the case of constant parameters) by carefully changing the path integral measure. In the process it becomes evident how to generalize the path integral approach to traditionally ``less-tractable" short-rate models such as the Black-Karasinski model, which generalization we discuss in Section \ref{sec.6}. We give explicit formulas for computing the bond pricing function in such models in the ``semiclassical" approximation. We make some concluding remarks in Section \ref{sec.7}, including an outline of how to apply perturbative techniques beyond the ``semiclassical" approximation. Appendix A provides some Gaussian path integrals. Appendix B gives some details for the change of measure.

\section{Classical Mechanics}\label{sec.2}

{}Newton's second law for one-dimensional motion along the $x$-axis reads:
\begin{equation}\label{N2}
 F = m~a
\end{equation}
where $m$ is the mass of an object, $a \equiv \ddot{x}$ is its acceleration,\footnote{\, Each dot over a variable stands for a time derivative.}  and $F$ is the force acting on it. Let
\begin{equation}
 F \equiv -{\partial V\over\partial x}
\end{equation}
where the function $V$ is called potential energy. Generally, $V$ is a function of $x$ and $t$. The equation of motion (\ref{N2}) then reads:
\begin{equation}\label{EOM}
 m\ddot{x} = -{\partial V\over\partial x}
\end{equation}
This equation can be derived via the principle of stationary action.\footnote{\, Often less precisely referred to as the principle of least action.} Let $x(t)$ be a continuous path ${\cal F}$ connecting two spacetime points $(x_0, t_0)$ and $(x_f, t_f)$, where $t_f > t_0$. The action functional $S \equiv S[x(t)]$ is defined as
\begin{equation}\label{action}
 S \equiv \int_{\cal F} dt~L(x, \dot{x}, t)
\end{equation}
where $L$ is the Lagrangian
\begin{equation}
 L \equiv {1 \over 2} m\dot{x}^2 - V
\end{equation}
Note that $L$ has explicit time dependence only if $V$ does. Now consider a small variation of the path $x(t) \rightarrow x(t) + \delta x(t)$, where $\delta x(t)$ vanishes at the endpoints of the path: $\delta x(t_0) = \delta x(t_f) = 0$. The variation of the action reads:
\begin{equation}
 \delta S = \int_{\cal F} dt~\left[{\partial L\over\partial x} - {d\over dt}{\partial L\over\partial \dot{x}}\right]\delta x(t)
\end{equation}
where we have integrated by parts and taken into account that the surface term vanishes. The functional derivative $\delta S/\delta x(t)$ vanishes if and only if \footnote{\, This is because $\delta x(t)$ is arbitrary subject to boundary conditions at $t_0$ and $t_f$.}
\begin{equation}\label{EL}
 {d\over dt}{\partial L\over\partial \dot{x}} = {\partial L\over\partial x}
\end{equation}
This is the Euler-Lagrange equation, which is the equation of motion (\ref{EOM}). So, the classical trajectory is determined by requiring that the action functional be stationary. Note that classical trajectories are deterministic: (\ref{EOM}) is a second order differential equation, so the path $x(t)$ is uniquely fixed by specifying the endpoints $(x_0, t_0)$ and $(x_f, t_f)$. Alternatively, it is uniquely determined by specifying the initial conditions: $x(t_0) = x_0$ and $\dot{x}(t_0) = v_0$, where $v_0$ is the velocity $v \equiv \dot{x}$ at $t = t_0$.

\section{Quantum Mechanics and Path Integral}\label{sec.3}

{}In contrast, quantum mechanics is not deterministic but probabilistic. One can only determine the probability $P(x_0, t_0; x_f, t_f)$ that starting at $(x_0, t_0)$ a quantum particle will end up at $(x_f, t_f)$. This is because of Heisenberg's uncertainty principle: in quantum mechanics it is not possible to specify both the position and the velocity at the same time with 100\% certainty (see below). One way to think about this is that, starting at $(x_0, t_0)$, the particle can take an infinite number of paths with a probability distribution. The probability $P(x_0, t_0; x_f, t_f)$ is given by
\begin{equation}
 P(x_0, t_0; x_f, t_f) = \left|\langle x_f, t_f\left.\right| x_0, t_0\rangle\right|^2
\end{equation}
where the probability amplitude\footnote{\, A.k.a. wave function, matrix element, propagator or correlator.} is given by Feynman's path integral (Feynman, 1948)\footnote{\, A.k.a. functional integral or infinite dimensional integral.}
\begin{equation}\label{FPI}
 \langle x_f, t_f\left.\right| x_0, t_0\rangle = \int_{x(t_0) = x_0,~x(t_f) = x_f} {\cal D}x~\exp\left({i\over\hbar}~S\right)
\end{equation}
where the integration is over all paths connecting points $(x_0, t_0)$ and $(x_f, t_f)$, and ${\cal D}x$ includes an appropriate integration measure, which we will define below. Also, $\hbar$ is the (reduced) Planck constant.

{}The path integral (\ref{FPI}) can be thought of as an $N\rightarrow \infty$ limit of $N-1$ integrals. Let us break up the interval $[t_0, t_f]$ into $N$ subintervals: $t_i \equiv t_{i-1} + \Delta t_i$, $i=1,\dots,N$, with $t_N \equiv t_f$. Let $x_i \equiv x(t_i)$, with $x_N\equiv x_f$. We can discretize the derivative $\dot{x}(t_i)$ via $(x(t_i) - x(t_{i-1}))/\Delta t_i = (x_i - x_{i-1})/\Delta t_i$, $i=1,\dots, N$. The integral in the action (\ref{action}) can be discretized as follows\footnote{\, As usual, there are choices in defining the discretized derivative and integral, which are essentially immaterial in the continuum limit.}
\begin{equation}
 S_{N} \equiv \sum_{i=1}^N \Delta t_i \left[{m(x_i - x_{i-1})^2\over 2\Delta t_i^2} - V(x_{i-1}, t_{i-1})\right]
\end{equation}
Then the path integral (\ref{FPI}) can be defined as
\begin{equation}\label{discrete}
 \int_{x(t_0) = x_0,~x(t_f) = x_f} {\cal D}x~\exp\left({iS\over\hbar}\right) \equiv \lim_{N \to \infty} \prod_{i=1}^{N} \sqrt{m\over 2\pi i \hbar\Delta t_i} \int \prod_{i=1}^{N-1} dx_i~\exp\left({iS_N\over\hbar}\right)
\end{equation}
where each of the $(N-1)$ integrals\footnote{\, We have $(N-1)$ integrals as $x_N$ is fixed: $x_N = x_f$.} over $x_1,\dots,x_{N-1}$ is over the real line ${\bf R}$. We will not derive the normalization of the measure in (\ref{discrete}). In the context of quantum mechanics it can be thought of being fixed either by using (\ref{discrete}) and (\ref{FPI}) as the {\em definition} of the probability amplitude and comparing it with the experiment, or by matching it to equivalent formulations of quantum mechanics, {\em e.g.}, Schr\"odinger's equation, which itself is compared with the experiment. In the context of stochastic processes, path integral interpretation is different than in quantum mechanics and we will fix the measure directly using the definition of (conditional) expectation.

\subsection{Classical Limit}

{}Considering path integral in quantum mechanics is useful as based on physical intuition it helps develop methods that are also applicable in finance. In quantum mechanics path integral provides an intuitive picture for making a connection with deterministic classical dynamics. Intuitively, quantum effects are associated with $\hbar$ being nonzero. Thus, according to Heisenberg's uncertainty principle
\begin{equation}
 \sigma_x~\sigma_p \geq {\hbar \over 2}
\end{equation}
where $\sigma_x$ and $\sigma_p$ are standard deviations of the position $x$ and momentum $p\equiv m\dot{x}$. So, $\hbar$ is the measure of deviation from classical dynamics -- in the limit $\hbar \to 0$ both position and momentum (or, equivalently, velocity $v = \dot{x}$) can be known, hence classical determinism. Path integral provides an elegant and intuitive way of understanding this. In the $\hbar\to 0$ limit, the exponential factor $\exp(iS/\hbar)$ in the path integral (\ref{FPI}) oscillates very rapidly, so the main contribution to the path integral comes form those paths that make the action stationary, and these are precisely the classical paths from the Euler-Lagrange equation (\ref{EL}). The classical trajectory dominates the path integral in the $\hbar\to 0$ limit.

\subsection{Semiclassical Approximation}

{}Since classical paths dominate in the small $\hbar$ limit, fluctuations around classical paths should describe quantum corrections. This simple observation makes path integral into a powerful computational tool. Let $x_{cl}(t)$ be a solution to the Euler-Lagrange equation (\ref{EL}) subject to the boundary conditions $x_{cl}(t_0) = x_0$ and $x_{cl}(t_f) = x_f$. Let $x(t)$ be a general path with the same boundary conditions: $x(t_0) = x_0$ and $x(t_f) = x_f$. Let $\xi(t) \equiv x(t) - x_{cl}(t)$. This quantum fluctuation vanishes at the endpoints of the path:
\begin{equation}\label{xi.b}
 \xi(t_0) = \xi(t_f) = 0
\end{equation}
Furthermore, we can decompose the action into the classical and quantum pieces:
\begin{equation}
 S = S_{cl} + S_{qu}
\end{equation}
where
\begin{equation}
 S_{cl} \equiv S[x_{cl}(t)] = \int_{t_0}^{t_f} dt~\left[{1\over 2}m \dot{x}_{cl}^2 - V(x_{cl}(t), t)\right]
\end{equation}
and
\begin{equation}
 S_{qu} \equiv \int_{t_0}^{t_f} dt~ L_{qu}(\xi, \dot{\xi}, t)
\end{equation}
where
\begin{equation}\label{L.qu}
 L_{qu}(\xi, \dot{\xi}, t) \equiv {1\over 2}m\dot{\xi}^2 - {1\over 2} \left.{\partial^2 V\over\partial x^2}\right|_{x = x_{cl}}\xi^2 - \sum_{k=3}^\infty {1\over k!}\left.{\partial^k V\over\partial x^k}\right|_{x = x_{cl}} \xi^k
\end{equation}
There is no linear term in $\xi$ in $S_{qu}$ as it vanishes due to the Euler-Lagrange equation (\ref{EL}) for the classical trajectory $x_{cl}(t)$ and the boundary conditions (\ref{xi.b}) for $\xi(t)$.

{}If we now define ${\widetilde \xi} \equiv \sqrt{\hbar}~\xi$, we have\footnote{\, Note that this is consistent with the measure (\ref{discrete}), which contains $1/\sqrt{\hbar}$ for each integration.}
\begin{equation}
 {S\over \hbar} = {S_{cl}\over \hbar} + S_{qu}^{(2)}[{\widetilde\xi}] + \sum_{k=3}^\infty \hbar^{(k - 1)/2}~S_{qu}^{(k)}[{\widetilde \xi}]
\end{equation}
where $S_{qu}^{(2)}$ corresponds to the quadratic in $\xi$ term in (\ref{L.qu}), while $S_{qu}^{(k)}$, $k > 2$ correspond to the higher order terms. The latter are suppressed by extra powers of $\sqrt{\hbar}$ and the leading quantum correction comes from the quadratic piece. Keeping only the quadratic piece is known as the semiclassical approximation.\footnote{\, A.k.a. the WKB (Wentzel-Kramers-Brillouin) approximation.} If $V$ is quadratic in $x$, then there are no higher order terms and this produces an exact result.\footnote{\, $V(x) = m\omega^2 x^2/2$ is the potential for a harmonic oscillator.} Here we will not do any computations in the case of quantum mechanics as we are interested in applying path integral to asset pricing. However, the physical picture has lead us to a computational tool, whereby we keep only quadratic terms in the action and treat higher order terms as perturbative -- in this case, quantum -- corrections. As we will see, this can be applied to asset pricing as well. This brings us to Euclidean path integral.

\subsection{Euclidean Path Integral}\label{sub.euclidean}

{}Mathematically, the complex phase in Feynman's path integral (\ref{FPI}) might be a bit unsettling.\footnote{\, While mathematically Feynman's path integral might not be strictly well-defined, Feynman diagram techniques based on it have been used to compute various quantities in Quantum Electrodynamics with mindboggling precision. {\em E.g.}, independent determinations of the fine structure constant experimentally agree within $10^{-8}$ precision.} The complex phase disappears if we go to the so-called Euclidean time via the Wick rotation $t\to -it$. We then have Euclidean path integral:
\begin{equation}\label{EPI}
 \langle x_f, t_f\left.\right| x_0, t_0\rangle = \int_{x(t_0) = x_0,~x(t_f) = x_f} {\cal D}x~\exp\left(-{S_E\over\hbar}\right)
\end{equation}
where
\begin{equation}
 S_E = \int dt~L_E(x,\dot{x}, t)
\end{equation}
and\footnote{\, It is assumed that, if there is any explicit $t$-dependence in $V$, it is such that $V$ is real.}
\begin{equation}
 L_E = {1\over 2} m\dot{x}^2 + V
\end{equation}
The Euler-Lagrange equation is still of the form (\ref{EL}) with $L$ replaced by $L_E$. In the remainder, we will focus on Euclidean path integral (relevant in asset pricing), so moving forward we will drop the subscript ``$E$".

{}Mathematically, Euclidean path integral looks more ``well-defined" than Feynman's path integral, at least for $V\geq 0$, as in this case the argument of the exponent is a real non-negative number. The discretized version is defined via
\begin{equation}\label{discrete.E}
 \int_{x(t_0) = x_0,~x(t_f) = x_f} {\cal D}x~\exp\left(-{S\over\hbar}\right) \equiv \lim_{N \to \infty} \prod_{i=1}^N \sqrt{m\over 2\pi \hbar\Delta t_i} \int \prod_{i=1}^{N-1} dx_i~\exp\left(-{S_{N}\over\hbar}\right)
\end{equation}
where
\begin{equation}
 S_{N} \equiv \sum_{i=1}^N \Delta t_i \left[{m(x_i - x_{i-1})^2\over 2\Delta t_i^2} + V(x_{i-1}, t_{i-1})\right]
\end{equation}
Here too we can consider fluctuations around classical paths $x(t) = x_{cl}(t) + \xi(t)$, and the semiclassical approximation amounts to keeping only the terms quadratic in $\xi$;
\begin{equation}
 S = S_{cl} + S_{qu}
\end{equation}
where
\begin{equation}
 S_{cl} \equiv S[x_{cl}(t)] = \int_{t_0}^{t_f} dt~\left[{1\over 2}m \dot{x}_{cl}^2 + V(x_{cl}(t), t)\right]
\end{equation}
and
\begin{equation}
 S_{qu} \equiv \int_{t_0}^{t_f} dt~ L_{qu}(\xi, \dot{\xi}, t)
\end{equation}
where
\begin{equation}\label{L.qu.E}
 L_{qu}(\xi, \dot{\xi}, t) \equiv {1\over 2}m\dot{\xi}^2 + {1\over 2} \left.{\partial^2 V\over\partial x^2}\right|_{x = x_{cl}}\xi^2 + \sum_{k=3}^\infty {1\over k!}\left.{\partial^k V\over\partial x^k}\right|_{x = x_{cl}} \xi^k
\end{equation}
Next we discuss how to compute the path integral in the semiclassical approximation.

\subsection{Gaussian Path Integral}\label{sub3.4}

{}Let us drop ${\cal O}(\xi^3)$ terms (if any) in (\ref{L.qu.E}). We then have
\begin{equation}\label{semi}
 \langle x_f, t_f\left.\right| x_0, t_0\rangle = \exp\left(-{S_{cl}\over\hbar}\right)\int_{{\widetilde\xi}(t_0)={\widetilde\xi}(t_f) = 0} {\cal D}\xi~\exp\left(-S^{(2)}[{\widetilde\xi}]\right)
\end{equation}
where ${\widetilde \xi}\equiv \sqrt{m/\hbar}~\xi$,
\begin{equation}\label{S.2}
 S^{(2)}[{\widetilde\xi}] \equiv {1\over 2} \int dt~{\widetilde\xi}(t)~C~{\widetilde\xi}(t)
\end{equation}
and
\begin{equation}\label{C}
 C\equiv -{d^2\over dt^2} + U(t)
\end{equation}
is a second-order differential operator.\footnote{\, Known as the Schr\"odinger operator.} Here
\begin{equation}
 U(t)\equiv {1\over m} \left.{\partial^2 V\over \partial x^2}\right|_{x = x_{cl}}
\end{equation}
and in rewriting $S^{(2)}[{\widetilde\xi}]$ via (\ref{S.2}) we have used the boundary conditions ${\widetilde\xi}(t_0)={\widetilde\xi}(t_f) = 0$. Also, we kept ${\cal D}\xi$ in (\ref{semi}) -- we have to fix the measure anyway.

{}So, we have a Schr\"odinger operator $C$ on the interval $t\in[t_0,t_f]$ with Dirichlet boundary conditions ${\widetilde\xi}(t_0)={\widetilde\xi}(t_f) = 0$. Let $\psi_n(t)$ be a complete orthonormal set of eigenfunctions of $C$ satisfying the boundary conditions $\psi_n(t_0) = \psi_n(t_f) = 0$:
\begin{eqnarray}
 && C~\psi_n(t) = \lambda_n~\psi_n(t)\\
 && \sum_n \psi_n(t)~\psi_n(t^\prime) = \delta(t-t^\prime)\\
 && \int_{t_0}^{t_f} \psi_n(t)~\psi_m(t) = \delta_{nm}
\end{eqnarray}
We have the following expansion:
\begin{equation}
 {\widetilde\xi}(t) = \sum_n c_n~\psi_n(t)
\end{equation}
and
\begin{equation}
 S^{(2)}[{\widetilde\xi}] = {1\over 2}\sum_n \lambda_n~c_n^2
\end{equation}
The integration measure ${\cal D}\xi$ can be written as
\begin{equation}
 {\cal D}\xi = {\cal N} \prod_n {dc_n\over\sqrt{2\pi}}
\end{equation}
where ${\cal N}$ is a normalization constant to be determined. The path integral in (\ref{semi}) then reads:
\begin{equation}
 \int_{{\widetilde\xi}(t_0)={\widetilde\xi}(t_f) = 0} {\cal D}\xi~\exp\left(-S^{(2)}[{\widetilde\xi}]\right) = {\cal N}\prod_n \lambda^{-1/2}_n = {\cal N}~
 \left[\det(C)\right]^{-1/2}
\end{equation}
where the determinant of $C$ is formally defined as the product of its eigenvalues.

\subsubsection{Gelfand-Yaglom Theorem}

{}Computing Schr\"odinger operator determinants is facilitated by the Gelfand-Yaglom theorem (Gelfand and Yaglom, 1960)\footnote{\, Also see, {\em e.g.}, (Burghelea {\em et al}, 1991), (Coleman, 1979), (Dunne, 2008), (Forman, 1987), (Kleinert and Chervyakov, 1998), (Kleinert, 2004), (Kirsten and McKane, 2003, 2004), (Levit and Smilansky, 1977), (Simon, 1977).} according to which
\begin{equation}
 {\det(C_1)\over\det(C_2)} = {\phi_1(t_f) \over \phi_2(t_f)}
\end{equation}
where $C_1$ and $C_2$ are two Schr\"odinger operators with Dirichlet boundary conditions at $t_0$ and $t_f$, and
\begin{eqnarray}
 &&C_1~\phi_1 = 0,~~~\phi_1(t_0) = 0,~~~\dot{\phi}_1(t_0) = 1\\
 &&C_2~\phi_2 = 0,~~~\phi_2(t_0) = 0,~~~\dot{\phi}_2(t_0) = 1
\end{eqnarray}
Applying this theorem to $C_1 = C$ and $C_2 = C_*$, where $C$ is defined in (\ref{C}), while $C_*$ is the Schr\"odinger operator for a free particle ($V=0$)
\begin{equation}
 C_* \equiv -{d^2\over dt^2}
\end{equation}
and noting that $\phi_*(t) = t - t_0$, we have
\begin{equation}
 \det(C) = \det(C_*)~{\phi(t_f) \over {t_f - t_0}}
\end{equation}
where $\phi(t)$ is the solution to the following initial value problem:
\begin{equation}\label{phi.eq}
 \left[-{d^2\over dt^2} + U(t)\right]\phi(t) = 0,~~~\phi(t_0) = 0,~~~\dot{\phi}(t_0) = 1
\end{equation}
which is straightforward to implement numerically, if need be (see below).

{}So, our path integral reduces to
\begin{equation}\label{cal.N}
 \int_{{\widetilde\xi}(t_0)={\widetilde\xi}(t_f) = 0} {\cal D}\xi~\exp\left(-S^{(2)}[{\widetilde\xi}]\right) = {\widetilde {\cal N}}~\sqrt{{t_f - t_i}\over \phi(t_f)}
\end{equation}
where
\begin{equation}
 {\widetilde {\cal N}} \equiv \left[\det(C_*)\right]^{-1/2}~{\cal N}
\end{equation}
We do not even need to compute ${\cal N}$ -- which can be done by discretizing, computing $(N-1)$ integrals and then taking $N\rightarrow\infty$ limit -- because if we know the amplitude $\langle x_f, t_f\left.\right| x_0, t_0\rangle_*$ for the free particle case, we can directly compute ${\widetilde {\cal N}}$ instead. This is precisely the approach we will follow in the context of asset pricing.

\subsubsection{Van Vleck-Pauli-Morette Formula}

{}In the cases where $S_{cl}$ as a function of $x_0$ and $x_f$ is explicitly known, we do not even need to solve (\ref{phi.eq}). We can apply the Van Vleck-Pauli-Morette Formula instead:
\begin{equation}
 {1\over\phi(t_f)} = - {\partial^2 S_{cl}\over\partial x_0\partial x_f}
\end{equation}
In practice, however, often it might be easier to solve (\ref{phi.eq}).

\subsection{Operator Expectation Values}

{}Thus far we have discussed the probability amplitude $\langle x_f, t_f\left.\right| x_0, t_0\rangle$. More generally, we can consider expectation values
\begin{equation}
 \langle x_f, t_f\left|A(t)\right| x_0, t_0\rangle \equiv \int_{x(t_0) = x_0,~x(t_f) = x_f} {\cal D}x~\exp\left(-{S\over \hbar}\right) A(t)
\end{equation}
where the l.h.s. is interpreted as an expectation value of an operator $A(t)$, while on the r.h.s. $A(t)$ is a function(al) constructed from $x(t)$, its derivatives and $t$. Let $A \equiv \exp(-{\widetilde A}/\hbar)$. Then we have
\begin{equation}
 \langle x_f, t_f\left|A(t)\right| x_0, t_0\rangle \equiv \int_{x(t_0) = x_0,~x(t_f) = x_f} {\cal D}x~\exp\left(-{{\widetilde S}\over\hbar}\right)
\end{equation}
where ${\widetilde S} \equiv S + {\widetilde A}$. In general, when ${\widetilde A}$ is not a local function of $t$ but a functional, approximating this path integral via a semiclassical approximation by expanding $S$ to the quadratic order in fluctuations around the classical solution to the Euler-Lagrange equation based on $S$ would be incorrect. Instead, we would have to expand around a classical solution to the Euler-Lagrange equation based on the ``effective" action ${\widetilde S}$. However, ${\widetilde S}$ may not even be local. In this case the methods discussed above cannot be straightforwardly applied. This will restrict ${\widetilde A}$ (see below).

\section{Path Integral in Asset Pricing}\label{sec.4}

{}Euclidean path integral naturally arises in asset pricing. Suppose we have a stock $S_t$ and a cash bond $B_t$. Let us assume that $B_t$ is deterministic. Suppose $X$ is a claim\footnote{\, {\em E.g.}, this could be a call/put/binary option or some other derivative.} at maturity $T$. Then the price of the claim at time $t$ is given by
\begin{equation}
 V_t = B_t\langle B_T^{-1}X\rangle_{{\bf Q},{\cal F}_t}
\end{equation}
Here ${\bf Q}$ is the measure under which the discounted stock price $Z_t\equiv B_t^{-1}S_t$ is a martingale, and the conditional expectation $\langle\cdot\rangle_{{\bf Q},{\cal F}_t}$ is defined along the latter portion of paths that have initial segments ${\cal F}_t$. Below we discuss such conditional expectations in the path integral language. Our discussion is general and not limited to stocks.

{}Consider a ${\bf P}$-Brownian motion\footnote{\, A.k.a. a Wiener process.} $W_t$ between $t=0$ and some horizon time $T$ (here ${\bf P}$ is the measure).
Let $x(t)$ be the values of $W_t$ (with $x(0)=0$). We will
divide the time interval $[t_0,t_f]$, $0\leq t_0<t_f\leq T$, into $N$ subintervals
$[t_{i-1},t_i]$, $t_N\equiv t_f$, $t_i-t_{i-1}\equiv \Delta t_i>0$. Let $x_i \equiv x(t_i)$, $x_N\equiv x_f$,
$\Delta x_i\equiv
x_i-x_{i-1}$. Let $A_t$, $0\leq t\leq T$, be a previsible process, {\em i.e.}, $A_t$ depends only on the path
${\cal F}_t=\{(x(s),s)|s\in [0,t]\}$:
\begin{equation}
 A_t=A({\cal F}_t)
\end{equation}
The conditional expectation (here ${\cal F}_{t_0}=\{(x_*(s),s)|s\in [0,t_0],
x_*(0)=0, x_*(t_0)=x_0\}$, where $x_*(s)$ is fixed)
\begin{equation}
 \langle A_{t_f} \rangle_{{\bf P},{\cal F}_{t_0}}
\end{equation}
can be thought of as a $\Delta t_i\rightarrow 0$,
{\em i.e.}, $N\rightarrow \infty$, limit of the corresponding discrete expression:\footnote{\, In this expectation only $x(t_0)=x_0$ is fixed and $x_N = x_f$ is integrated over -- see below.}
\begin{equation}\label{discrete.Br}
 \langle A_{t_f} \rangle_{{\bf P},{\cal F}_{t_0}}=
 \lim_{N\to\infty}~\prod_{i=1}^N \int_{-\infty}^\infty {dx_i\over
 \sqrt{2\pi \Delta t_i}}\exp\left(-{(\Delta x_i)^2\over 2\Delta t_i}\right)~
 A_{t_f,{\cal F}_{t_0}}
\end{equation}
where
\begin{equation}
 A_{t_f,{\cal F}_{t_0}}=A({\cal F}_{t_0}\cup \{(x_1,t_1),
 \dots,(x_N,t_N)\})
\end{equation}
This limit is nothing but a Euclidean path integral
\begin{equation}\label{BrPI}
 \langle A_{t_f} \rangle_{{\bf P},{\cal F}_{t_0}}=
 \int_{x(t_0) = x_0} {\cal D}x~\exp(-S)~A_{t_f,{\cal F}_{t_0}}
\end{equation}
where ${\cal D}x$ includes a properly normalized measure (see below), and
\begin{equation}
 S[x] \equiv \int_{t_0}^{t_f} {{\dot x}^2(t)\over 2}~dt
\end{equation}
is the Euclidean action functional
for a free particle on ${\bf R}$ (as before, dot in ${\dot x}(t)$ denotes time derivative).

{}Let us note some straightforward differences in units between the path integral in (\ref{BrPI}) and the quantum mechanical Euclidean path integral (\ref{EPI}). Here we have no mass $m$ or $\hbar$, and $x(t)$ does not have the dimension of length but of $\sqrt{t}$. {\em I.e.}, the two path integrals are the same in the units where $m=\hbar=1$. In these units the discretized measures in (\ref{discrete.E}) and (\ref{discrete.Br}) are identical. Note that the measure in (\ref{discrete.Br}) is a corollary of the measure ${\bf P}$ for the Brownian motion.\footnote{\, Recall that the measure (\ref{discrete.E}) is fixed by requiring agreement with experiment, be it directly or via its derivation using, {\em e.g.}, Schr\"odinger's equation, which itself is verified experimentally.}

{}In the context of quantum mechanics we considered paths where both endpoints are fixed. Here too we have the following conditional expectation:
\begin{eqnarray}
 \langle A_{t_f} \rangle_{{\bf P},{\cal F}_{t_0},x(t_f) = x_f} &=&
 \lim_{N\to\infty}~ \prod_{i=1}^{N-1} \int_{-\infty}^\infty dx_i \prod_{i=1}^N {1\over
 \sqrt{2\pi \Delta t_i}}\exp\left(-{(\Delta x_i)^2\over 2\Delta t_i}\right)~
 A_{t_f,{\cal F}_{t_0}} \nonumber\\
 &=&\int_{x(t_0) = x_0,~x(t_f) = x_f} {\cal D}x~\exp(-S)~A_{t_f,{\cal F}_{t_0}}\label{discrete.Br.1}
\end{eqnarray}
Note that
\begin{equation}
 \langle A_{t_f} \rangle_{{\bf P},{\cal F}_{t_0}} = \int_{-\infty}^\infty dx^\prime~\langle A_{t_f} \rangle_{{\bf P},{\cal F}_{t_0},x(t_f) = x^\prime}
\end{equation}
In the asset pricing context we will primarily be interested in the latter conditional expectation.

{}For $A_t \equiv 1$ the conditional expectation (\ref{discrete.Br.1}) is nothing but the probability\footnote{\, Note that in the quantum mechanical context this quantity has the interpretation of the probability {\em amplitude} instead.} of starting at $(x_0,t_0)$ and ending at $(x_f,t_f)$:
\begin{equation}
 \langle 1 \rangle_{{\bf P},{\cal F}_{t_0},x(t_f) = x_f}  = P(x_0,t_0; x_f, t_f) = {1\over\sqrt{2\pi(t_f-t_0)}}\exp\left(-{(x_f-x_0)^2\over 2(t_f-t_0)}\right)
\end{equation}
The discussion in Subsection \ref{sub3.4} carries over unchanged (with $m=\hbar=1$), and we can immediately fix ${\widetilde {\cal N}}$ in (\ref{cal.N}) by noting that $S_{cl} = (x_f-x_0)^2/2(t_f-t_0)$ and $\phi(t) = t- t_0$ in this case, so that:
\begin{equation}
 {\widetilde {\cal N}} = {1\over\sqrt{2\pi(t_f-t_0)}}
\end{equation}
The same result can be obtained by doing a direct (and tedious, albeit straightforward) computation using the discretized definition (\ref{discrete.Br.1}) and taking the large $N$ limit. Using the path integral techniques gets us to the answer quickly and elegantly.

{}Now that we have fixed ${\widetilde {\cal N}}$, we can compute (the analog of) the ``semiclassical" approximation\footnote{\, Or the WKB approximation in the PDE language. We will use ``semiclassical" approximation.} for expectations (\ref{discrete.Br.1}) with $A_t$ of the form
\begin{eqnarray}\label{exp.A}
 && A_t = \exp\left(-\int_0^t dt^\prime \left[\rho(x(t^\prime), t^\prime)~\dot{x}(t^\prime) + V(x(t^\prime), t^\prime)\right]\right) \equiv A_{t_0}~\exp(-{\widetilde A}_t)
\end{eqnarray}
where $\rho(x,t)$ and $V(x,t)$ are deterministic functions. Thus, let
\begin{equation}
 {\widetilde S} \equiv S + {\widetilde A}_t
\end{equation}
Then the ``semiclassical" approximation is given by
\begin{equation}\label{semi.A}
 \langle A_{t_f} \rangle_{{\bf P},{\cal F}_{t_0},x(t_f) = x_f} = {A_{t_0}\over\sqrt{2\pi\phi(t_f)}}~\exp\left(-{\widetilde S}_{cl}\right)
\end{equation}
where ${\widetilde S}_{cl} \equiv {\widetilde S}[x_{cl}]$, $x_{cl}(t)$ is the solution of the Euler-Lagrange equation (subject to the boundary conditions $x_{cl}(t_0) = x_0$ and $x_{cl}(t_f) = x_f$)
\begin{equation}
 0 = {d\over dt}{\partial {\widetilde L} \over \partial \dot{x}} - {\partial{\widetilde L}\over \partial x} = \ddot{x} + {\partial\rho(x,t)\over\partial t} - {\partial V(x,t)\over \partial x}
\end{equation}
where the ``effective" Lagrangian ${\widetilde L}$ is defined via
\begin{equation}
 {\widetilde S} = \int_{t_0}^{t_f} dt~{\widetilde L}
\end{equation}
and is given by
\begin{equation}
 {\widetilde L} = {1\over 2}~\dot{x}^2 + \rho(x, t)~\dot{x}(t) + V(x, t)
\end{equation}
Also,
\begin{eqnarray}\label{phi}
 &&-\ddot{\phi}(t) + U(t)\phi(t) = 0,~~~\phi(t_0) = 0,~~~\dot{\phi}(t_0) = 1\\
 &&U(t)\equiv {\partial^2 V(x_{cl}, t)\over \partial x^2} - {\partial^2 \rho(x_{cl}, t)\over\partial_x\partial_t}
\end{eqnarray}
Note that locally the $\rho(x,t)~\dot{x}$ term in ${\widetilde L}$ simply shifts $\partial V/\partial x$ by $\partial\rho/\partial t$.

{}We discuss some explicit examples in Appendix A. Here the following remark is in order. The choice of the form of $A_t$ in (\ref{exp.A}) is based on the requirement that the ``effective" action ${\widetilde S}$ be a local functional containing no higher-than-then-first derivative of $x$. If ${\widetilde L}$, even if it is local, depends on higher derivatives of $x(t)$, then we no longer have a Schr\"odinger operator in the approximation where we neglect higher-than-quadratic terms in $x$ in ${\widetilde S}$. If ${\widetilde L}$ contains nonlocal terms, {\em e.g.},
\begin{equation}
 {\widetilde L}(x, \dot{x}, t) = {1\over 2}~\dot{x}^2 + {1\over 2}~\left[\int_{t_a}^{t_b} dt^\prime~\gamma(t^\prime) ~x(t^\prime)\right]^2
\end{equation}
where $\gamma(s)$ is some deterministic function (and, {\em e.g.}, $t_a = 0$ and $t_b = t$, or $t_a = t_0$ and $t_b = t_f$),
then the methods we discuss cannot be straightforwardly applied.\footnote{\, Nonlocal ${\widetilde L}$ can be thought of as an infinite series of terms containing higher derivatives of $x$.}

\section{Application to Short-rate Models}\label{sec.5}

{}The path integral methods can be applied to pricing bonds in short-rate models. A short-rate model posits a risk-neutral measure {\bf Q} and a short-rate process $r_t$. The cash bond process is given by
\begin{equation}
 B_t=\exp\left(\int_0^t r_s~ds\right)
\end{equation}
while the bond price is given by
\begin{equation}
 P(t,T)=\left\langle\exp\left(-\int_t^T r_s~ds\right)\right\rangle_{{\bf Q}, {\cal F}_t}
\end{equation}
The price at time $t$ of a general claim $X$ at maturity $T$ is
\begin{equation}
 v_t=\left\langle\exp\left(-\int_t^T r_s~ds\right)X\right\rangle_{{\bf Q}, {\cal F}_t}
\end{equation}
Consider the claim $X=1$ (to avoid confusion with the potential $V(x,t)$, we use lower case $v$ for the pricing function):
\begin{equation}\label{pricing.function}
 v(z,t,T)\equiv
 \left\langle\exp\left(-\int_t^T r_s~ds\right)\right\rangle_{{\bf Q},r_t=z}
\end{equation}
for which $v(r_t,t,T)=P(t,T)$, and $v(z,T,T)=1$.

{}In short-rate models one usually works with a parameterized family of processes, and chooses the parameters to best fit the market. Thus, let us {\em assume} that $r_t$ satisfies the following SDE:
\begin{equation}\label{r.markov}
 dr_t=\sigma(r_t,t) dW_t+\nu(r_t,t) dt
\end{equation}
where $\sigma(y,t)$ and $\nu(y,t)$ are deterministic functions, and $W_t$ is a ${\bf Q}$-Brownian motion. The pricing function $v(z, t, T)$ satisfies a pricing PDE, which follows from the requirement that the discounted bond process
$Z(t,T)\equiv B_t^{-1}P(t,T)=B_t^{-1}v(r_t,t,T)$ be a martingale under the risk-neutral measure ${\bf Q}$:
\begin{equation}\label{PDE}
 \nu(r_t,t) \partial_z v(r_t,t,T)+ \partial_t
 v(r_t,t,T)+{1\over 2}\sigma^2(r_t,t) \partial^2_z v(r_t,t,T)-r_t v(r_t,t, T)=0
\end{equation}
with the boundary condition $v(z,T,T)=1$.

\subsection{Path Integral Approach}

{}Instead of assuming (\ref{r.markov}) and solving the pricing PDE, following the previous section, we can write $v(z,t,T)$ as a path integral. We will do this in two steps. We will first discuss the Vasicek/Hull-White model. Then we will discuss a generalization.

{}In the Vasicek/Hull-White model, the short-rate SDE reads:
\begin{equation}
 dr_t = \sigma(t)dW_t + \left[\theta(t) - \alpha(t)r_t\right]dt
\end{equation}
where $\sigma(t)$, $\theta(t)$ and $\alpha(t)$ depend only on time. For constant $\sigma$, $\theta$ and $\alpha$ we have the mean-reverting Ornstein-Uhlenbeck process. For now, we will not assume that $\sigma(t)$, $\theta(t)$ and $\alpha(t)$ are constant.

{}Let
\begin{eqnarray}
 &&\beta(t,T)\equiv\exp\left(-\int_t^T\alpha(s)ds\right)\\
 &&\eta(t,T)\equiv\int_t^T\beta(t,u)du\label{eta}
\end{eqnarray}
Straightforward algebra yields
\begin{equation}
 \int_t^T r_s~ds = r_t\eta(t,T) + \int_t^T\eta(s, T)\left[\sigma(s)dW_s + \theta(s)ds\right]
\end{equation}
Our pricing function $v(z,t,T)$ therefore reads:
\begin{eqnarray}
 &&v(z,t,T) = \left\langle\exp\left(-r_t\eta(t,T) - \int_t^T\eta(s, T)\left[\sigma(s)dW_s + \theta(s)ds\right]\right)\right\rangle_{{\bf Q},r_t=z}=\nonumber\\
 &&\,\,\,\,\,\,\,\exp\left(-z\eta(t,T) - \int_t^T ds \left[\eta(s, T)\theta(s) - {1\over 2}~\eta^2(s, T)\sigma^2(s)\right]\right)\label{VHW}
\end{eqnarray}
where we have used (\ref{exp.lin}). Note that this result is exact.

\subsection{Direct Path Integral Computation}

{}One shortcoming of the Vasicek/Hull-White model is that $r_t$ can occasionally become negative. One way to deal with this is to consider short-rate models of the form $r_t = r_0 f(X_t)/f(X_0)$, where $X_t$ follows the Vasicek/Hull-White model:
\begin{equation}
 dX_t = \sigma(t)dW_t +\left[\theta(t) - \alpha(t)X_t\right]dt
\end{equation}
Here $f(y)$ is a positive function, {\em e.g.}, $f(y) = \exp(y)$, which is the Black-Karasinski model. From the path integral we can immediately see that in the general case the ``effective" action would be nonlocal. In the case where $\sigma$, $\theta$ and $\alpha$ are constant, the problem is tractable.\footnote{\, More precisely, the problem is tractable even if $\theta$ and $\alpha$ are $t$-dependent -- see below.}

{}We will now derive the result (\ref{VHW}) for constant $\sigma$, $\theta$ and $\alpha$ via a direct path integral computation using a change of measure.\footnote{\, A similar, but more ``heuristic", computation was performed in (Otto, 1998).} In the process it will become clear how to tackle more general cases (including the Black-Karasinski model). We have the ${\bf Q}$-Brownian motion $W_t$. Consider the following ${\bf P}$-Brownian motion:
\begin{equation}
 {\widetilde W}_t \equiv W_t + \int_0^t \gamma_s~ds
\end{equation}
where the previsible process $\gamma_t$ is given by (note that $r_t = X_t$ in this case)
\begin{equation}
 \gamma_t \equiv {{\theta - \alpha~r_t}\over\sigma}
\end{equation}
The change of measure is given by\footnote{\, Pursuant to the Cameron-Martin-Girsanov theorem, for which path integral provides an elegant proof -- see (Kakushadze, 2015) for details.}
\begin{equation}\label{measure.PQ}
 {d{\bf Q}\over d{\bf P}} = \exp\left(\int_0^T \gamma_s~d{\widetilde W}_s - {1\over 2}\int_0^T\gamma^2_s~ds\right)
\end{equation}
Furthermore, we have
\begin{equation}
 dr_t = \sigma~d{\widetilde W}_t
\end{equation}
and
\begin{eqnarray}
 &&r_t = r_0 + \sigma~{\widetilde W}_t\\
 &&\gamma_t = \nu - \alpha~{\widetilde W}_t\\
 &&\nu\equiv {{\theta - \alpha~r_0}\over\sigma}
\end{eqnarray}
The pricing function is given by
\begin{equation}
 v(z, t, T) = \left\langle\exp\left(\int_t^T\left[\gamma_s~d{\widetilde W}_s - {1\over 2}~\gamma^2_s~ds - r_s~ds\right]\right)\right\rangle_{{\bf P},{\widetilde W}_t=(z - r_0)/\sigma}\label{r.exp.p}
\end{equation}
This is a Gaussian path integral -- recall that ${\widetilde W}_s$ is replaced by $x(s)$ and $d{\widetilde W}_s$ is replaced by $\dot{x}(s)~ds$. We can simplify the computation by observing that for constant parameters the pricing function depends on $t$ and $T$ only in the combination $T-t$, so it suffices to compute $v(z, 0, T)$, where we have $r_0 = z$, so $\nu = (\theta - \alpha z)/\sigma$ and the initial condition on $x$ is $x(0) = 0$. Yet another simplification is achieved by changing integration from $x$ to $y\equiv x - \nu/\alpha$, so that $\gamma_s$ is replaced by $-\alpha y(s)$ (and the measure is not affected). We then have the following path integral:
\begin{eqnarray}
 &&v(z,0,T) = {\cal G}~\int_{-\infty}^\infty dy^\prime~\exp\left(-{\theta\over\alpha}~T- {\alpha\over 2}\left[\left(y^\prime\right)^2 - {\nu^2\over\alpha^2}\right] \right)\times\nonumber\\
 &&\,\,\,\,\,\,\,\times\int_{y(0) = -\nu/\alpha,~y(T) = y^\prime} {\cal D}y~\exp\left(-{\widetilde S}\right)
\end{eqnarray}
where the ``effective" action ${\widetilde S} = {\widetilde S}[y(s)]$ is given by
\begin{equation}
 {\widetilde S} = \int_0^T ds~{\widetilde L}(y(s),\dot{y}(s))
\end{equation}
and the ``effective" Lagrangian reads:
\begin{equation}
 {\widetilde L}(y(s), \dot{y}(s)) = {1\over 2}~\dot{y}^2(s) + V(y(s))
\end{equation}
where
\begin{equation}\label{V.VHW}
 V(y(s)) = {1\over 2}~\alpha^2 y^2(s)+ \sigma y(s)
\end{equation}
Also, ${\cal G}$ is a normalization factor. We have accounted for the change of measure, so it might seem that ${\cal G}$ should be 1. However, this is not so. We will come back to ${\cal G}$ momentarily, after we evaluate our path integral.

{}The Gaussian path integral over ${\cal D}y$ can be done using (\ref{HO1}). A tedious but straightforward computation gives
\begin{eqnarray}
 &&v(z,0,T) = {\cal G}~\exp\left(-{1\over 2}~\alpha T\right)\times\nonumber\\
 &&\,\,\,\,\,\,\,\times\exp\left(-z{\widetilde\eta}(T) - {\theta\over\alpha}\left[T - {\widetilde\eta}(T)\right] + {\sigma^2\over 2\alpha^2}\left[T-{\widetilde\eta}(T) - {\alpha\over 2}~{\widetilde\eta}^2(T)\right]\right)\\
 &&{\widetilde\eta}(T)\equiv {{1 - \exp(-\alpha T)}\over \alpha}\label{VHW1}
\end{eqnarray}
This precisely agrees with (\ref{VHW}) with constant coefficients except for the prefactor ${\cal G}\exp(-\alpha T/2)$. This is because
\begin{equation}\label{cal.G}
 {\cal G} = \exp\left({1\over 2}~\alpha T\right)
\end{equation}
To see this, let us compute the expectation
\begin{equation}
u(z,0, T)\equiv\left\langle\exp\left(\int_0^T\left[\gamma_s~d{\widetilde W}_s - {1\over 2}~\gamma^2_s~ds\right]\right)\right\rangle_{{\bf P},{\widetilde W}_t=0}
\end{equation}
This should be identically equal to 1. Let us do the path integral calculation as above:
\begin{eqnarray}\label{u}
 &&u(z, 0, T) = {\cal G}~\int_{-\infty}^\infty dy^\prime~\exp\left(- {\alpha\over 2}\left[\left(y^\prime\right)^2 - {\nu^2\over\alpha^2}\right] \right)\times\nonumber\\
 &&\,\,\,\,\,\,\,\times\int_{y(0) = -\nu/\alpha,~y(T) = y^\prime} {\cal D}y~\exp\left(-{\widehat S}\right)
\end{eqnarray}
where the ``effective" action ${\widehat S} = {\widehat S}[y(s)]$ is given by
\begin{equation}
 {\widehat S} = \int_0^T ds~\left[{1\over 2}~\dot{y}^2(s) + {1\over 2}~\alpha^2 y^2(s)\right]
\end{equation}
This gives
\begin{equation}
 u(z, 0, T) = {\cal G}~\exp\left(-{1\over 2}~\alpha T\right)
\end{equation}
So, we indeed have (\ref{cal.G}). This additional normalization factor arises due to the integration of $y^\prime = y(T)$ in (\ref{u}): we must integrate over the variable such that it is equivalent to the integration over $x^\prime = x(T)$ under the ${\bf Q}$ measure, and this variable is $y(T)~\exp(\alpha T/2)$. We discuss this in more detail in Appendix \ref{app.B}.\footnote{\, For a related discussion in the discretized picture, see, {\em e.g.}, (Moriconi, 2004).}

\section{Generalization to Positive Short-rate Models}\label{sec.6}

{}It is now clear how to generalize the path integral approach to general models of the form
\begin{eqnarray}
 &&r_t = r_0~{f(X_t, t)\over f(X_0, 0)}\\
 &&dX_t = \sigma~dW_t +\left[\theta - \alpha~X_t\right]dt\label{X.sigma}
\end{eqnarray}
with constant $\sigma$, $\theta$ and $\alpha$. (We will discuss $t$-dependent coefficients below.) Without loss of generality we can set $X_0 = 0$ and $f(0,0)=1$.

{}We have the ${\bf Q}$-Brownian motion $W_t$. Consider the following ${\bf P}$-Brownian motion:
\begin{equation}
 {\widetilde W}_t \equiv W_t + \int_0^t \gamma_s~ds
\end{equation}
where the previsible process $\gamma_t$ is given by
\begin{equation}
 \gamma_t \equiv {{\theta - \alpha~X_t}\over\sigma}
\end{equation}
The change of measure is given by (\ref{measure.PQ}). We have
\begin{equation}
 dX_t = \sigma~d{\widetilde W}_t
\end{equation}
and
\begin{eqnarray}\label{X.W}
 &&X_t = \sigma~{\widetilde W}_t\\
 &&\gamma_t = \nu - \alpha~{\widetilde W}_t\\
 &&\nu\equiv {{\theta}\over\sigma}
\end{eqnarray}
The pricing function is given by
\begin{equation}
 v(z, t, T) = \left\langle\exp\left(\int_t^T\left[\gamma_s~d{\widetilde W}_s - {1\over 2}~\gamma^2_s~ds - r_0~f(\sigma {\widetilde W}_s, s)~ds\right]\right)\right\rangle_{{\bf P},{\widetilde W}_t=x_*}\label{r.exp.p.1}
\end{equation}
where $x_*$ is determined from the equation
\begin{equation}\label{x.star}
 r_0~f(\sigma x_*, t) = z
\end{equation}
We can write the pricing function $v(x, t, T)$ as a path integral: ${\widetilde W}_s$ is replaced by $x(s)$ and $d{\widetilde W}_s$ is replaced by $\dot{x}(s)~ds$. A simplification is achieved by changing integration from $x$ to $y\equiv x - \nu/\alpha$, so that $\gamma_s$ is replaced by $-\alpha y(s)$ (and the measure is not affected). We then have the following path integral:\footnote{\, As we are allowing explicit time-dependence in the function $f(X_t, t)$, the pricing function no longer depends only on the $T-t$ combination.}
\begin{eqnarray}
 &&v(z,t,T) = {\cal G}~\int_{-\infty}^\infty dy^\prime~\exp\left(- {\alpha\over 2}\left[\left(y^\prime\right)^2 - y_*^2\right] \right)\times\nonumber\\
 &&\,\,\,\,\,\,\,\times\int_{y(t) = y_*,~y(T) = y^\prime} {\cal D}y~\exp\left(-{\widetilde S}\right)\label{path.f}
\end{eqnarray}
where
\begin{equation}
 y_* \equiv x_* - {\theta\over\sigma\alpha}
\end{equation}
The ``effective" action ${\widetilde S} = {\widetilde S}[y(s)]$ is given by
\begin{equation}\label{eff.action.bk}
 {\widetilde S} = \int_t^T ds~{\widetilde L}(y(s),\dot{y}(s), s)
\end{equation}
and the ``effective" Lagrangian reads:
\begin{equation}
 {\widetilde L}(y(s), \dot{y}(s), s) = {1\over 2}~\dot{y}^2(s) + V(y(s), s)
\end{equation}
where
\begin{equation}
 V(y(s), s) = {1\over 2}~\alpha^2 y^2(s)+ r_0~f\left(\sigma y(s) + {\theta\over\alpha}, s\right)
\end{equation}
Also, ${\cal G}$ is a normalization factor. It is fixed from the requirement that
\begin{eqnarray}
 &&1 = {\cal G}~\int_{-\infty}^\infty dy^\prime~\exp\left(- {\alpha\over 2}\left[\left(y^\prime\right)^2 - y_*^2\right] \right)\times\nonumber\\
 &&\,\,\,\,\,\,\,\times\int_{y(t) = y_*,~y(T) = y^\prime} {\cal D}y~\exp\left(-{\widehat S}\right)
\end{eqnarray}
where
\begin{equation}
 {\widehat S} = \int_0^T ds~\left[{1\over 2}~\dot{y}^2(s) + {1\over 2}~\alpha^2 y^2(s)\right]
\end{equation}
This gives us
\begin{equation}\label{cal.G.1}
 {\cal G} = \exp\left({1\over 2}~\alpha(T-t)\right)
\end{equation}
in complete parallel with the previous section -- see Appendix \ref{app.B} for details.

{}The path integral over ${\cal D}y$ in (\ref{path.f}) is not Gaussian for a general function $f(X_t, t)$. We can use the ``semiclassical" approximation:
\begin{equation}\label{price.gen}
 v(z,t,T) = \exp\left({1\over 2}~\alpha(T-t)\right)~\int_{-\infty}^\infty dy^\prime~{1\over \sqrt{2\pi\phi(T)}}~\exp\left(- {\alpha\over 2}\left[\left(y^\prime\right)^2 - y_*^2\right] -{\widetilde S}_{cl}\right)
\end{equation}
where ${\widetilde S}_{cl} \equiv {\widetilde S}[y_{cl}(s)]$, and $y_{cl}(s)$ is determined from
\begin{equation}
 \ddot{y}_{cl}(s) = {\partial V(y_{cl}(s), s)\over \partial y},~~~y_{cl}(t) = y_*,~~~y_{cl}(T) = y^\prime
\end{equation}
Also, $\phi(T)$ is given by
\begin{eqnarray}
 &&-\ddot{\phi}(s) + U(s)\phi(s) = 0,~~~\phi(t) = 0,~~~\dot{\phi}(t) = 1\\
 &&U(s)\equiv {\partial^2 V(y_{cl}(s), s)\over \partial y^2}
\end{eqnarray}
Note that (\ref{price.gen}) is exact if the function $f(X_t,t)$ is quadratic (or linear).

\subsection{Black-Karasinski Model}

{}Let us illustrate the above discussion on the Black-Karasinski model:
\begin{equation}
 f(X_t, t) = \exp(X_t)
\end{equation}
We then have
\begin{eqnarray}
 &&y_* = {1\over\sigma}~\ln\left(z\over r_*\right)\\
 &&r_* \equiv r_0~\exp\left(\theta\over\alpha\right)\\
 &&V(y) = {1\over 2}~\alpha^2 y^2 + r_*^2~\exp(\sigma y)\\
 &&U(s) = \alpha^2 + r_*\sigma^2~\exp(\sigma y_{cl}(s))
\end{eqnarray}
where $y_{cl}(s)$ satisfies the following equation of motion
\begin{equation}
 \dot{y}^2_{cl}(s) - \alpha^2 y_{cl}^2(s) - 2r_*\exp\left(\sigma y_{cl}(s)\right) = 2E,~~~y(t) = y_*,~~~y(T) = y^\prime
\end{equation}
where $E$ is an integration constant. Computing $v(z,t,T)$ is now straightforward: $y_{cl}$ and $\phi(T)$ are obtained by solving differential equations, and the integral over $y^\prime$ is fast-converging.

\subsection{``Quadratic" Model}

{}In fact, to ensure that the short-rate is nonnegative, we do not even need a highly nonlinear model such as the Black-Karasinski model. Thus, consider a model with
\begin{equation}\label{quad}
 f(X_t, t) = 1 + b(t) X_t + a(t) X_t^2
\end{equation}
So long as $b^2(t) < 4a(t)$ and $a(t) > 0$, the short-rate is strictly positive.\footnote{\, If we wish to allow zero short-rate, we can relax this condition to $b^2(t)\leq 4a(t)$.} The ``semiclassical" approximation is exact -- the path integral is Gaussian -- even for non-constant $a(t)$ and $b(t)$. For the sake of simplicity, let us focus on the case with constant coefficients $a$ and $b$. There is a key difference between the quadratic model (\ref{quad}) and the linear case. Thus, from (\ref{VHW1}) (taking into account (\ref{cal.G})) it follows that the asymptotic ({\em i.e.}, large $T$) yield in the Vasicek/Hull-White model (with constant coefficients) is negative unless $\sigma^2 \leq 2\alpha\theta$. In the path integral language, this is due to the unbounded from below linear term (corresponding to volatility) in the potential (\ref{V.VHW}), which competes with the quadratic term (corresponding to mean-reversion). For $\sigma^2 > 2\alpha\theta$ volatility ``wins" and modes corresponding to negative values of the short-rate contribute into the bond prices.\footnote{\, The spectrum -- or, roughly, the set of short-rate values contributing to bond and other claim prices -- is discrete in the Vasicek/Hull-White model. However, it is nonnegative only if $\sigma^2 \leq 2\alpha\theta$.} In contrast, in the quadratic model (\ref{quad}) the asymptotic yield is always positive, irrespective of $\sigma$. In the path integral language, this is due to the additional quadratic term in the potential (\ref{V.VHW}) (stemming from the quadratic term in (\ref{quad})), which grows with volatility and compensates for the negative contribution from the linear term.\footnote{\, If $b^2 \leq 4a$, then the spectrum in the quadratic model is strictly positive, irrespective of $\sigma$. In fact, to obtain nonnegative spectrum this condition can be further relaxed (see below).}

{}The quadratic model (\ref{quad}) is a good illustration of the usefulness of the path integral approach. In the path integral language things are simple -- we just need to understand how to compute Gaussian path integrals once and use the same tool in various models over and over again. However, in the PDE language things look substantially more complicated. Thus, the short-rate SDE reads:
\begin{eqnarray}
 &&{dr_t\over r_0} = 2\sqrt{a}\sigma\sqrt{{r_t\over r_0} - \left(1 - {b^2\over 4a}\right)}dW_t + \left[2\alpha\left(1 - {b^2\over 4a}\right) + a\sigma^2 +\right.\nonumber\\
 &&\,\,\,\,\,\,\,+\left.2\sqrt{a}\left(\theta + {\alpha b\over 2a}\right)
 \sqrt{{r_t\over r_0} - \left(1 - {b^2\over 4a}\right)} - 2\alpha{r_t\over r_0}\right]dt\label{quad.SDE}
\end{eqnarray}
Let us define
\begin{equation}
 {\widetilde W_t} \equiv W_t + {1\over \sigma}\left(\theta + {\alpha b\over 2a}\right)t
\end{equation}
Then we have the following SDE:
\begin{equation}
 {dr_t\over r_0} = 2\sqrt{a}\sigma\sqrt{{r_t\over r_0} - \left(1 - {b^2\over 4a}\right)}d{\widetilde W}_t + \left[2\alpha\left(1 - {b^2\over 4a}\right) + a\sigma^2 - 2\alpha{r_t\over r_0}\right]dt\label{quad.SDE.1}
\end{equation}
which is a shifted Cox-Ingersoll-Ross process. However, while $W_t$ is a ${\bf Q}$-Brownian motion, where ${\bf Q}$ is the risk-neutral measure, ${\widetilde W}_t$ generally\footnote{\, In the special case $b = -2a\theta/\alpha$ (for which we must have $a<\alpha^2/\theta^2$) we have ${\widetilde W}_t = W_t$.} is not a ${\bf Q}$-Brownian motion -- it is a ${\bf P}$-Brownian motion under a different measure ${\bf P}$. So, nothing is gained by rewriting the SDE (\ref{quad.SDE}) via (\ref{quad.SDE.1}). We would still need to solve a rather complicated-looking pricing PDE (\ref{PDE}) corresponding to (\ref{quad.SDE}), which we will not spell out here for the sake of brevity. It would appear unlikely to write down the SDE (\ref{quad.SDE}) and the corresponding pricing PDE based on simple intuitive arguments. All the while, in the path integral language this model is exactly solvable.\footnote{\, The path integral computation in the quadratic model is straightforward and can be performed similarly to that in the Vasicek/Hull-White model using the Gaussian path integral formulas in Appendix \ref{app.A}. We skip details for the sake of brevity, especially in the context of the next subsection.}

\subsection{Short-rate as Potential}

{}As mentioned in Introduction, path integral in finance is neither a panacea, nor is it intended to yield ``fundamentally new results". However, it does provide some advantages in some cases. Thus, as we argued in the previous subsection, in the quadratic model, in some sense, it offers a ``computational" advantage over a bruit-force numerical approach to the pricing PDE (\ref{PDE}) as the path integral in this case is Gaussian, which can be evaluated analytically. In this case, it also provides an ``intuitional" advantage not only because of the analytical solution, but also because the path integral approach makes writing down this model evident, while in the PDE language the same model appears to be very convoluted and not as ``intuitive".

{}Here we give another example of the path integral language providing a simple intuitive picture that allows to reach qualitatively nontrivial conclusions and come up with fresh ideas. Thus, let us go back to the pricing function (\ref{pricing.function}). Instead of assuming the SDE (\ref{r.markov}) or that $r_t$ is a function of $X_t$, where $X_t$ follows (\ref{X.sigma}), an alternative approach is to consider short-rate processes of the form $r_t = V(W_t, t)$, where the function $V(x, t)$ is such that $r_t$ has the desirable properties, {\em e.g.}, $V(x,t)$ is bounded from below (or both from below and above). Then (\ref{pricing.function}) is given by the following path integral\footnote{\, The initial condition $V(x(t),t) = z$ may not have a unique solution for $x(t)$. Then we must sum over the contributions corresponding to all such solutions.}
\begin{equation}
 v(x,t,T) = \int_{V(x(t),t)=z} {\cal D}x~\exp(-{\widetilde S})
\end{equation}
where the ``effective" action reads:
\begin{equation}
 {\widetilde S} = \int_t^T ds \left[{\dot{x}^2(s)\over 2} + V(x(s), s)\right]
\end{equation}
This is nothing but the path integral for a Euclidean particle with the potential $V(x,t)$. So, in this framework the short-rate plays the role of the potential. Therefore, without doing any computations, we can conclude that, in time-homogeneous cases where the potential $V(x)$ has no explicit time dependence, in order to have a nonnegative asymptotic bond yield, it is not required that $V(x,t)$, {\em i.e.}, the short-rate $r_t$, be nonnegative. It suffices that $V(x,t)$ be such that the energy spectrum is discrete and the ground state energy is nonnegative. {\em E.g.}, consider a harmonic oscillator potential $V(x) = {1\over 2}\omega^2 x^2 - V_0$. The energy spectrum is given by $E_n = \left(n+ {1\over 2}\right)\omega - V_0$, $n=0,1,2,\dots$, so the ground state energy\footnote{\, In quantum mechanics $E_n = \left(n+ {1\over 2}\right)\hbar\omega - V_0$, so $E_0 > -V_0$ is a purely quantum effect.} $E_0\geq 0$ provided that $V_0\leq {1\over 2}\omega$, and the short-rate need not be positive but is bounded by $r_t\geq -\omega/2$. This opens a new avenue in simple short-rate models such as the Ho and Lee model, where the short-rate is allowed to be negative, the discrete spectrum is achieved by introducing a reflecting barrier for the underlying Brownian motion $W_t$ (as opposed to $r_t$), and the resulting bond yield curve has a sensible shape with a positive asymptotic yield. Furthermore, this allows analytical treatment of the cases with more realistic time-dependent drift.\footnote{\, Time-dependent drift is less tractable when the reflecting barrier is for $r_t$ instead of $W_t$.}  This idea was recently explored in (Kakushadze, 2016).

{}To summarize, the path integral in asset pricing -- just as in physics -- is a complementary approach. Thus, in physics, there are problems where path integral is easier, {\em e.g}, perturbation theory via Feynman diagrams (see below). But there are problems where it is not, {\em e.g.}, solving the hydrogen atom spectrum using path integral would be cumbersome and the Schr\"odinger equation is a more convenient approach. Similarly, in the asset pricing context, {\em e.g.}, pricing bonds in the quadratic model (\ref{quad}) appears to be much easier using path integral than by solving the pricing PDE (\ref{PDE}). However, solving problems with barriers (as in (Kakushadze, 2016)), where one deals with boundaries, typically is easier in the PDE language.

\section{Concluding Remarks}\label{sec.7}

{}In our discussion above, it was important to have (\ref{X.W}) in the sense that it allows us to determine $x_*$ in (\ref{x.star}). If $x_*$ cannot be determined, than we would be stuck. It is a separate issue that if $\sigma$ is $t$-dependent, the change in the path integration measure is more nontrivial and requires special treatment (see, {\em e.g.}, (Otto, 1998)). However, this is a moot point as without being able to fix $x_*$ we cannot do the path integral anyway. So, we will assume that $\sigma$ is constant. However, {\em a priori}, there is no obstruction to having $t$-dependent $\theta$ and $\alpha$. Things are a bit more complicated, but still tractable. All the necessary ingredients are provided above, so the reader should be able to work out the details.

{}Another point concerns the validity of the ``semiclassical" approximation. In quantum mechanics it is the leading quantum correction -- the higher corrections corresponding to higher orders in quantum fluctuations $\xi$ over the classical background $x_{cl}$ are suppressed in the small $\hbar$ limit. In the asset pricing context we have no $\hbar$. The analog of ``quantumness" is randomness or volatility -- if the volatility is zero, there is no randomness. So, the meaning of the ``semiclassical" approximation is that it is a small ``volatility" approximation. However, this is not to say that this is a small-$\sigma$ approximation, where $\sigma$ is defined in (\ref{X.sigma}). Indeed, $\sigma$ is a dimensionful parameter. The dimensionless expansion parameter is $\epsilon \equiv \sqrt{T-t}~\sigma$, and the ``semiclassical" approximation is valid when $\epsilon\ll 1$. This can be seen by rescaling $s\equiv (T-t)~{\widetilde s}$ and $y\equiv \sqrt{T-t}~{\widetilde y}$ in the ``effective" action (\ref{eff.action.bk}).

{}Higher corrections beyond the ``semiclassical" approximation can be computed using perturbation theory. The basic idea is that we can evaluate the path integral
\begin{equation}\label{path.pert}
 \int_{\xi(t_0)=\xi(t_f) = 0} {\cal D}\xi~\exp\left(-\int_{t_0}^{t_f} dt~L_{qu}(\xi, \dot{\xi}, t)\right)\equiv K_{qu}(x_0, t_0; x_f, t_f)
\end{equation}
where
\begin{equation}\label{L.qu.1}
 L_{qu}(\xi, \dot{\xi}, t) \equiv {1\over 2}\dot{\xi}^2 + {1\over 2} \left.{\partial^2 V\over\partial x^2}\right|_{x = x_{cl}}\xi^2 + \sum_{k=3}^\infty {1\over k!}\left.{\partial^k V\over\partial x^k}\right|_{x = x_{cl}} \xi^k\equiv L^{(2)}_{qu}(\xi, \dot{\xi}, t) + {\widehat V}_{qu}\left(\xi, t\right)
\end{equation}
by expanding the exponent so we have an infinite series of (integrals over) correlators of the form
\begin{eqnarray}
 &&\langle\xi(\tau_1)\xi(\tau_2)\dots\xi(\tau_n)\rangle \equiv \nonumber\\
 &&\,\,\,\,\,\,\,\int_{\xi(t_0)=\xi(t_f) = 0} {\cal D}\xi~\exp\left(-\int_{t_0}^{t_f} dt~L^{(2)}_{qu}(\xi, \dot{\xi}, t)\right) \xi(\tau_1)\xi(\tau_2)\dots\xi(\tau_n)
\end{eqnarray}
with $n\geq 3$, where $L^{(2)}_{qu}(\xi, \dot{\xi}, t)$ is the quadratic part in (\ref{L.qu.1}). These $n$-point correlators can be evaluated using the generating functional
\begin{equation}
 Z[J] \equiv  \int_{\xi(t_0)=\xi(t_f) = 0} {\cal D}\xi~\exp\left(-\int_{t_0}^{t_f} dt \left[L^{(2)}_{qu}(\xi, \dot{\xi}, t)-J(t)\xi(t)\right]\right)
\end{equation}
which is a Gaussian path integral and can be readily computed. Then we have
\begin{equation}
 \langle\xi(\tau_1)\xi(\tau_2)\dots\xi(\tau_n)\rangle = \left.{1\over Z[0]}~{\delta\over\delta J(\tau_1)}{\delta\over\delta J(\tau_2)}\dots{\delta\over\delta J(\tau_n)}~Z[J]\right|_{J=0}
\end{equation}
and
\begin{equation}\label{path.pert.1}
 K_{qu}(x_0, t_0; x_f, t_f) = \left.\exp\left[-\int_{t_0}^{t_f} dt~{\widehat V}_{qu}\left({\delta\over\delta J(t)}, t\right)\right] Z[J]\right|_{J=0}
\end{equation}
where ${\widehat V}_{qu}\left(\xi, t\right)$ contains cubic and/or higher terms in $\xi$.
The correlator (\ref{path.pert.1}) can then be computed order-by-order in perturbation theory. At each order taking the functional derivatives becomes a combinatorial problem, which was solved by Feynman via a neat diagrammatic representation in terms of Feynman diagrams constructed from propagators and interaction vertices (see, {\em e.g.}, (Corradini, 2014), (Kleinert, 2004) and (Rattazzi, 2009)) which can be readily used in asset pricing.

\appendix
\section{Conditional Expectations via Path Integral}\label{app.A}

{}Here we give examples of conditional expectations for processes of the form (\ref{exp.A}) discussed in Section \ref{sec.4} via path integral. The expressions below are exact, {\em i.e.}, the ``semiclassical" approximation is exact as these path integrals are Gaussian.

{}$\bullet$ $V(x, t)\equiv 0$, $\rho(x, t) \equiv \rho(t)$ is independent of $x$. We have
\begin{eqnarray}
 &&\left\langle \exp\left(-\int_0^{t_f} dt~\rho(t)~\dot{x}(t)\right) \right\rangle_{{\bf P},{\cal F}_{t_0},x(t_f) = x_f} = \exp\left(-\int_0^{t_0} dt~ \rho(t)~\dot{x}(t)\right)\times\nonumber\\
 &&\,\,\,\,\,\,\, {1\over\sqrt{2\pi(t_f-t_0)}}~\exp\left(-{\left[x_f - x_0 - \int_{t_0}^{t_f} dt~\rho(t)\right]^2\over 2(t_f-t_0)} + {1\over 2}\int_{t_0}^{t_f} dt~\rho^2(t)\right)
\end{eqnarray}
\begin{eqnarray}
 &&\left\langle \exp\left(-\int_0^{t_f} dt~\rho(t)~\dot{x}(t)\right) \right\rangle_{{\bf P},{\cal F}_{t_0}} = \exp\left(-\int_0^{t_0} dt~ \rho(t)~\dot{x}(t)\right)\times\nonumber\\
 &&\,\,\,\,\,\,\, \exp\left( {1\over 2}\int_{t_0}^{t_f} dt~\rho^2(t)\right)\label{exp.lin}
\end{eqnarray}
The last equation reproduces a well-known result.

{}$\bullet$ $V(x, t) = {1\over 2}\omega^2 x^2$, $\rho(x, t)\equiv 0$, $\omega = \mbox{const}$.
Eq. (\ref{phi}) reads
\begin{equation}
 -\ddot{\phi}(t) + \omega^2 \phi(t) = 0,~~~\phi(t_0) = 0,~~~\dot{\phi}(t_0) = 1
\end{equation}
so we have $\phi(t) = \sinh(\omega (t-t_0))/\omega$. Also,
\begin{eqnarray}
 &&x_{cl}(t) = {{x_0~\sinh\left[\omega(t_f - t)\right] + x_f~\sinh\left[\omega(t - t_0)\right]}\over\sinh\left[\omega(t_f - t_0)\right]}\\
 &&{\widetilde S}_{cl} = {\omega\over 2}~{{(x_0^2 + x_f^2)\cosh\left[\omega(t_f - t_0)\right] - 2x_0x_f}\over\sinh\left[\omega(t_f - t_0)\right]}
\end{eqnarray}
For the expectations, we have
\begin{eqnarray}\label{HO1}
 &&\left\langle \exp\left(-{\omega^2\over 2}\int_0^{t_f} dt~x^2(t)\right) \right\rangle_{{\bf P},{\cal F}_{t_0},x(t_f) = x_f} = \exp\left(-{\omega^2\over 2}\int_0^{t_0} dt~ x^2(t)\right)\times\nonumber\\
 &&\sqrt{\omega\over 2\pi\sinh\left[\omega(t_f - t_0)\right]}~\exp\left(-{\omega\over 2}~{{(x_0^2 + x_f^2)\cosh\left[\omega(t_f - t_0)\right] - 2x_0x_f}\over\sinh\left[\omega(t_f - t_0)\right]}\right)
\end{eqnarray}
\begin{eqnarray}
 &&\left\langle \exp\left(-{\omega^2\over 2}\int_0^{t_f} dt~x^2(t)\right) \right\rangle_{{\bf P},{\cal F}_{t_0}} = \exp\left(-{\omega^2\over 2}\int_0^{t_0} dt~ x^2(t)\right)\times\nonumber\\
 &&{1\over\sqrt{\cosh\left[\omega(t_f - t_0)\right]}}~\exp\left(-{\omega x_0^2\over 2}~\tanh\left[\omega(t_f - t_0)\right]\right)\label{HO2}
\end{eqnarray}
These results are obtained rather ``effortlessly" using path integral.

{}$\bullet$ $V(x,t ) = {1\over 2}\omega^2 x^2$, $\rho(x, t)\equiv \rho(t)$ is independent of $x$, $\omega = \mbox{const}$.
The $\rho(t)$ term does not affect $\phi(t)$ in this case, it only modifies ${\widetilde S}_{cl}$ via $x_{cl}$, which is the solution of the Euler-Lagrange equation of motion
\begin{equation}
 \ddot{x}_{cl} + \dot{\rho} = \omega^2~x_{cl}
\end{equation}
subject to the boundary conditions $x_{cl}(t_0) = x_0$, $x_{cl}(t_f) = x_f$ (we skip details for the sake of brevity).

\section{Change of Measure}\label{app.B}

{}Here we discuss the additional normalization factor ${\cal G}$ in (\ref{cal.G}) and (\ref{cal.G.1}). Let us start with the ${\bf Q}$-Brownian motion $W_t$ and the ${\bf P}$-Brownian motion ${\widetilde W}_t$ related via
\begin{equation}
 {\widetilde W}_t = W_t + \int_0^t \gamma_s~ds
\end{equation}
where the previsible process $\gamma_s$ is given by
\begin{equation}
 \gamma_s = -\alpha~{\widetilde W}_s
\end{equation}
The change of measure is given by (\ref{measure.PQ}). Let
\begin{equation}
 P(x_0,t_0; x_f, t_f) \equiv \langle 1\rangle_{W_{t_0} = x_0,W_{t_f} = x_f} = {1\over\sqrt{2\pi(t_f- t_0)}}~\exp\left(-{(x_f-x_0)^2\over 2(t_f-t_0)}\right)
\end{equation}
We can write this probability via path integral ($W_s$ is substituted by $x(s)$):
\begin{equation}
 P(x_0,t_0; x_f, t_f) = \int_{x(t_0)= x_0,~x(t_f)=x_f}{\cal D}x~\exp\left(-{1\over 2}\int_{t_0}^{t_f} \dot{x}^2(s)~ds\right)
\end{equation}
and we have
\begin{equation}
 \int_{-\infty}^\infty dx^\prime \int_{x(t_0)= x_0,~x(t_f)=x^\prime}{\cal D}x~\exp\left(-{1\over 2}\int_{t_0}^{t_f} \dot{x}^2(s)~ds\right) = 1
\end{equation}
Let us now go through the same steps upon changing the measure from ${\bf Q}$ to ${\bf P}$:
\begin{eqnarray}
 &&{\widetilde P}(y_0,t_0; y_f, t_f) \equiv \left\langle \exp\left(\int_{t_0}^{t_f}\left[\gamma_s~d{\widetilde W}_s - {1\over 2}~\gamma_s^2~ds\right]\right) \right\rangle_{{\widetilde W}_{t_0} = y_0,{\widetilde W}_{t_f} = y_f} = \nonumber\\
 &&\,\,\,\,\,\,\,\int_{y(t_0)= y_0,~y(t_f)=y_f}{\cal D}y~\exp\left(-{1\over 2}\int_{t_0}^{t_f} \left[\dot{y}(s) + \alpha y(s)\right]^2~ds\right)=\nonumber\\
 &&\,\,\,\,\,\,\,\sqrt{\alpha\over2\pi\sinh\left[\alpha(t_f-t_0)\right]}~\exp\left(-{{\alpha\left[{\widehat y}_f - {\widehat y}_0\right]^2}\over{2\sinh\left[\alpha(t_f-t_0)\right]}}\right)
\end{eqnarray}
where ${\widehat y}_f \equiv y_f\exp\left(\alpha (t_f - t_0)/2\right)$ and ${\widehat y}_0 \equiv y_0\exp\left(-\alpha (t_f - t_0)/2\right)$.
This is why under the measure ${\bf P}$ it is integrating over $y^\prime\equiv y_f\exp(\alpha(t_f-f_0)/2)$ (and not over $y^\prime\equiv y_f$) that is equivalent to integrating over $x^\prime\equiv x_f$ under the measure ${\bf Q}$, hence the extra normalization factor ${\cal G}$ in (\ref{cal.G}) and (\ref{cal.G.1}).

\subsection*{Acknowledgments}
{}I would like to thank Olindo Corradini for a discussion on path integral. I am grateful to Peter Carr, whose invitation to give a seminar at his group at Morgan Stanley motivated this write-up.

\end{document}